\documentclass{article}
\usepackage{spconf,amsmath,graphicx}
\usepackage{amsmath,graphicx,bbm}
\usepackage{multirow, hhline}

\title{Character-Aware Attention-Based End-to-End Speech Recognition}
\name{Zhong Meng, Yashesh Gaur, Jinyu Li, Yifan Gong}
\address{Microsoft Corporation, Redmond, WA, USA}

\begin{document}
\maketitle

\begin{abstract}

Predicting words and subword units (WSUs) as the output has shown to be effective for the attention-based encoder-decoder (AED) model in end-to-end speech recognition. However, as one input to the decoder recurrent neural network (RNN), each WSU embedding is learned independently  through context and acoustic information in a purely data-driven fashion. Little effort has been made to explicitly model the morphological relationships among WSUs. In this work, we propose a novel character-aware (CA) AED model in which each WSU  embedding is computed by summarizing the embeddings of its constituent characters using a CA-RNN. This WSU-independent CA-RNN is jointly trained with the encoder, the decoder and the attention network of a conventional AED to predict WSUs. With CA-AED, the embeddings of morphologically similar WSUs are naturally and directly correlated through the CA-RNN in addition to the semantic and acoustic relations modeled by a traditional AED. Moreover, CA-AED significantly reduces the model parameters in a traditional AED by replacing the large pool of WSU embeddings with a much smaller set of character embeddings. On a 3400 hours Microsoft Cortana dataset, CA-AED achieves up to 11.9\% relative WER improvement over a strong AED baseline with 27.1\% fewer model parameters.

\end{abstract}

\begin{keywords}
character-aware, end-to-end, attention, encoder-decoder, speech recognition
\end{keywords}

\section{Introduction}
Traditional hybrid automatic speech recognition (ASR) system \cite{sainath2011making, jaitly2012application, DNN4ASR-hinton2012,
deng2013recent} consists of an acoustic
model, a pronunciation model and a language model. Different components
are optimized separately towards different objectives.  With the advance of
deep learning, end-to-end (E2E) speech recognition has shown promising ASR
performance by incorporating the three components into a single deep neural
network (DNN) and directly mapping a sequence of input speech signal to a
sequence of output labels as the transcription. Connectionist temporal classification (CTC) \cite{graves2006connectionist, graves2014towards},
recurrent neural network transducer \cite{graves2012sequence} and
attention-based encoder-decoder (AED) \cite{cho2014properties, bahdanau2014neural, chorowski2015attention} are three dominant approaches that
enable E2E speech recognition. With the advantage of no conditional independence assumption over CTC, AED was first introduced to the speech area in \cite{chorowski2015attention}
for phoneme recognition. In AED model, an encoder maps the input speech frames into
high-level representations and a decoder predicts the current output symbol
given the acoustic context vector and the embeddings of previously predicted symbols. 
An attention mechanism
\cite{bahdanau2014neural} aligns each decoder output with the encoded
representations and computes the acoustic context vector. In \cite{bahdanau2016end,
chan2016listen}, AED is successfully applied to large vocabulary speech
recognition and is recently reported to achieve superior performance to
the conventional hybrid systems 
\cite{chiu2018state}.

Initially, characters (graphemes) are commonly used as the output units for
AED in E2E ASR \cite{bahdanau2016end, chan2016listen, lu2016training}. Later on, people began to
use words and subword units (WSUs) as the output since the perplexity
of a word LM is lower than that of a character LM and the WSUs enable a
stronger LM to be learned in the decoder of AED \cite{kannan2018analysis}. Modeling WSUs instead of
characters enables the E2E system to more directly target on the ASR output -- word hypotheses. 
One popular type of WSUs 
is the word pieces model generated by iteratively combining two units out of
the current inventory that increase the likelihood the most on the
training data \cite{chiu2018state, schuster2012japanese}.
Another kind of WSUs is the mixed-units \cite{li2018advancing} which include
all the frequent words in the vocabulary as the major part and decompose each infrequent word into frequent
words and leftover multi-character units.
Mixed units were first
introduced to address the issue of out-of-vocabulary (OOV) words
\cite{li2018advancing} in a CTC-based E2E system. Recently, for AED-based ASR, mixed units outperform the characters and words as the output units
\cite{gaur2019acoustic}. With around 30k WSUs commonly used for US English, the 
WSU set is about 1000 times larger than the character set (about 30). Therefore,
the WSU-based AED necessitates a much larger output layer with much more
parameters but requires fewer decoding steps to generate the ASR results.

The WSU-based AED model learns a distinct embedding vector for each WSU
from the text history and the speech signal by conditioning the decoder on
previous WSU embeddings to predict the current WSU posteriors.  Although
good performance is achieved, the morphological relationships among the WSUs are not
explicitly modeled or well exploited. In many languages, the semantic
relations of WSUs are not only determined by their relative positions and
functionality in the sentences, but also are directly reflected in the
similarity among their spellings, i.e., the shared characters that
form the WSUs. 

To directly capture the additional morphological relationships among WSUs,
we propose a character-aware (CA) AED in which only the character embeddings
are learned through the E2E training and each WSU representation is
generated by summarizing the embeddings of its constituent characters using
a CA-recurrent neural network (RNN). With CA-AED, the embeddings of different
WSUs that share the same character sequence are naturally bridged
through the WSU-independent CA-RNN. A rare WSU representation can be
better estimated through ``assembling'' the well-trained character
embeddings.  With the same output layer predicting WSU posteriors, 
CA-AED inherits the strong WSU discriminability in a large vocabulary and
further improves AED through more sophisticated character-aware modeling of WSU embeddings. 

Moreover, CA-AED significantly reduces the number of model parameters by
replacing a large pool of WSU embeddings with a much smaller
set of character embeddings. Therefore, CA-AED is expected to
outperform conventional AED models with remarkably reduced model size
and computational cost. A similar CA architecture based on convolutional
neural network was proposed to improve the perplexity in neural
language model \cite{kim2016character} and has outperformed the
word/morpheme-level long short-term memory network language model with fewer
parameters.

Evaluated on 3400 hours Microsoft Cortana dataset (US English) with models of different sizes, the proposed CA-AED achieves up to a 11.9\% relative word error rate (WER) improvement over a strong AED baseline with 27.1\% fewer model parameters for word-piece output, and up to 8.5\% relative WER gain with 29.3\% fewer parameters for mixed-unit output. 


\section{Attention-Based Encoder-Decoder (AED) Model for E2E ASR}
\label{sec:aed}
\begin{figure}[htpb!]
	\centering
	\includegraphics[width=0.85\columnwidth]{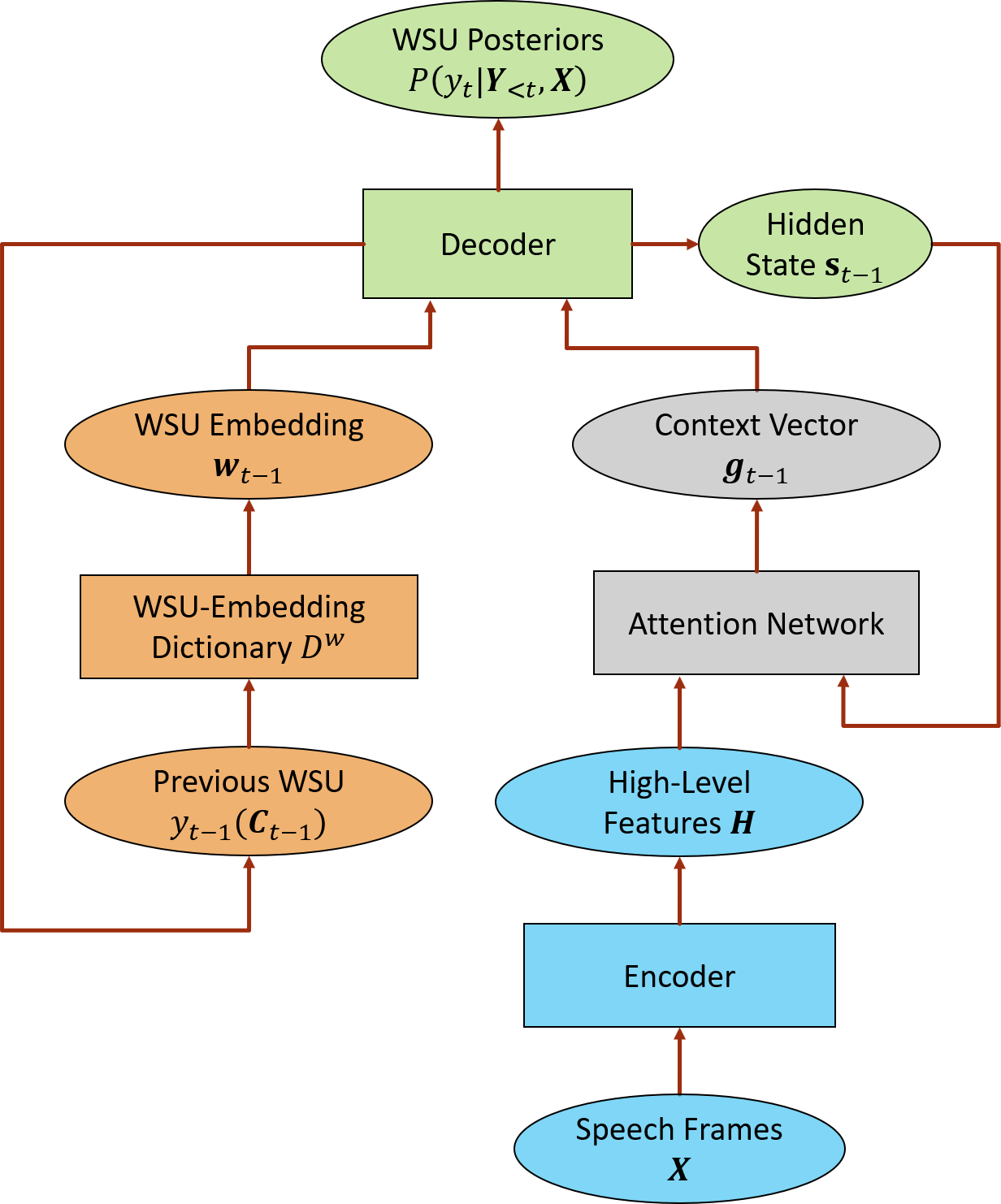}
	\caption{The architecture of AED model for E2E ASR. The
	convolution network generating vector $\mathbf{f}_{t, i}$ is
        omitted for brevity.}
	\label{fig:aed}
\end{figure}

In this work, we focus on improving the AED-based E2E speech recognition
\cite{chorowski2015attention, bahdanau2016end, chan2016listen} with WSUs as
the output units. AED models the conditional probability distribution $P(\mathbf{Y} |
\mathbf{X})$ over sequences of output WSU labels $\mathbf{Y}=\{y_1, \ldots,
y_T\}$ given a sequence of input speech frames $\mathbf{X}=\{\mathbf{x}_1,
\ldots, \mathbf{x}_I\}$, where $y_t \in \mathbbm{R}, t = 1, \ldots, T, \mathbf{x}_i
\in \mathbbm{R}^{d_x}, i = 1, \ldots, I$.  To achieve E2E ASR, AED directly maps
$\mathbf{X}$ to $\mathbf{Y}$ via an encoder, a decoder, an
attention network and a WSU-embedding dictionary as shown in Fig.
\ref{fig:aed}. 

The encoder is an
RNN which encodes the sequence of input speech frames $\mathbf{X}$
into a sequence of high-level features $\mathbf{H} =
\{\mathbf{h}_1, \ldots, \mathbf{h}_I\}$ as follows and it resembles
the role of an acoustic model in a traditional ASR system.
\begin{align}
	\mathbf{h}_i = \text{RNN}^{\text{enc}}(\mathbf{h}_{i-1},
	\mathbf{x}_i)
\end{align}
where $\mathbf{h}_i \in \mathbbm{R}^{d_h}$ represents the hidden state of
the encoder RNN at current time $i$. With the encoder,
$P(\mathbf{Y}|\mathbf{X})$ is equivalent to the probability over the output
WSU sequences conditioned on the encoded high-level features $\mathbf{H}$,
i.e., $P(\mathbf{Y}|\mathbf{H})$, as follows.
\begin{align}
	P(\mathbf{Y}|\mathbf{X}) = P(\mathbf{Y}|\mathbf{H}) =
	\prod_{t=1}^T P(y_t | y_0, \ldots, y_{t-1},
	\mathbf{H}) 
\end{align}

We use a decoder to model $P(\mathbf{Y}|\mathbf{H})$.  In
$P(y_t | y_0, \ldots,$ $y_{t-1}, \mathbf{H})$, the
conditional dependence of $y_t$ on $\mathbf{H}$ is captured
through an acoustic context vector $\mathbf{g}_t \in \mathbbm{R}^{d_h}$
obtained by a linear combination of all the encoded features $\mathbf{H}$
weighted by an attention probability vector $\mathbf{a}_t \in
\mathbbm{R}^I$ against $\mathbf{H}$. To estimate $\mathbf{a}_t$, a
location-aware attention mechanism \cite{chorowski2015attention} is applied
to determine which encoded features in $\mathbf{H}$ should the decoder
attend to predict the output label $y_t$. 
Specifically, $\mathbf{a}_t$ is computed by normalizing the similarity
scores, $z_{t,i}, i = 1, \ldots, I$, among the current hidden state
$\mathbf{s}_t \in \mathbbm{R}^{d_s}$ of the decoder RNN, each encoded
feature $\mathbf{h}_i$ and the convolved attention vector 
$\mathbf{f}_{t,i}$ as follows. 
\begin{align}
	z_{t,i} & = \mathbf{v}^{\top}\text{ReLU}(W_h
		\mathbf{h}_{i}
	+ W_s \mathbf{s}_t + W_f \mathbf{f}_{t,i} + \mathbf{b}_z), \footnotemark \\
	\mathbf{a}_t & = \text{softmax}(\mathbf{z}_{t}), \\
	\mathbf{g}_{t} & = \sum_{i = 1}^{I} a_{t,i}
		\mathbf{h}_{i},
	\label{eqn:att}
\end{align}
where the column vector $\mathbf{v}\in \mathbbm{R}^k$, bias $\mathbf{b}_z\in
\mathbbm{R}^k$, the projection matrices $W_h \in \mathbbm{R}^{k\times d_h}$, $W_s \in
\mathbbm{R}^{k\times d_s}$, $W_f \in \mathbbm{R}^{k\times d_f}$ are all learnable
parameters. $\mathbf{f}_{t,i} \in \mathbbm{R}^{d_f}$ is generated
by convolving the previous attention probability vector $\mathbf{a}_{t - 1}$ with
a matrix $F\in \mathbbm{R}^{d_f\times r}$. 
\footnotetext{We use ReLU instead tanh as non-linear activation function because it gives better ASR performance.}

The conditional dependence of $y_t$ on $y_{0}, \ldots, y_{t-1}$ is
modeled by an RNN with a feedback connection from the decoder output of the
previous time step to the input of the current step.  Similar to an RNN
language model \cite{mikolov2010recurrent}, we maintain a large dictionary $\mathcal{D}^w$
which maps each WSU to an embedding vector and feed the previous WSU
embedding instead of the label to the current input of the decoder. The WSU
embeddings are learned jointly with the other parts of the AED in the
training process. We denote the WSU-embedding sequence of $\mathbf{Y}$ as
$\mathbf{W}=\{\mathbf{w}_1, \ldots, \mathbf{w}_T\}$.
Therefore, at each
time step $t$, the decoder RNN takes the sum of the previous WSU embedding 
$\mathbf{w}_{t-1}$ and the acoustic context vector $\mathbf{g}_{t-1}$ as the
input to predict the conditional probability of each WSU, i.e., $P(u |
y_0, \ldots, y_{t-1}, \mathbf{H}), u \in \mathbbm{U}$, at the current time $t$ as follows, where
$\mathbbm{U}$ is the set of all the WSUs:
\begin{align}
       & \mathbf{s}_t = \text{RNN}^{\text{dec}}(\mathbf{s}_{t-1}, \mathbf{w}_{t-1} + 
       \mathbf{g}_{t-1}), \footnotemark \label{eqn:decoder_rnn} \\
       & \left[P(u | y_0, \ldots, y_{t-1}, \mathbf{H})\right]_{u \in
       \mathbbm{U}} = \nonumber \\
       & \qquad \quad \qquad \qquad \qquad \text{softmax}\left[W_{y}(\mathbf{s}_t + \mathbf{g}_t) +
       \mathbf{b}_y\right], \label{eqn:decoder_fc}
\end{align}
\footnotetext{In Eq.~\eqref{eqn:decoder_rnn} and Eq.~\eqref{eqn:decoder_fc}, we sum together the $\mathbf{g}_t$ and $\mathbf{s}_t$ (or $\mathbf{w}_t$) instead of concatenation, because, by summation, we get a lower-dimensional combined vector than concatenation, saving the number of parameters by half for the subsequent projection operation (i.e., half the size of $W_y$). In our experiments, concatenation does not improve the performance even with more parameters.}
where bias $\mathbf{b}_y\in \mathbbm{R}^k$ and the matrix $W_y \in
\mathbbm{R}^{d_y\times d_s}$ are learnable parameters. Note that $d_y$ is
the number of WSUs in the vocabulary and $d_h = d_s$ in our AED model.

To train the AED model, we maximize the conditional probability of the
reference label sequences $\mathcal{Y}=\{\mathbf{Y}_1, \ldots,
\mathbf{Y}_N\}$ given their corresponding input speech sequences
$\mathcal{X}=\{\mathbf{X}_1, \ldots, \mathbf{X}_N\}$ on the training
corpus, which is equivalent to minimizing the total cross-entropy loss
$\mathcal{L}_{\text{CE}}$ between the output of the decoder and the
references at all the time steps below:
\begin{align}
	\mathcal{L}_{\text{CE}} &= -\sum_{n = 1}^N \log P(\mathbf{Y}_n |
	\mathbf{X}_n) \nonumber \\
	& =-\sum_{n = 1}^N \sum_{t = 1}^{T_n} \log P(y_t^{(n)} | y_0^{(n)}, \ldots, y_{t-1}^{(n)},
       \mathbf{H}_n).
       \label{eqn:loss}
\end{align}

\section{Character-Aware (CA) AED Model for E2E ASR}
\label{sec:ca_aed}
As discussed in Section \ref{sec:aed}, a dictionary of WSU embeddings are
learned through the E2E training of the AED model. 
The WSU embeddings exhibit the property that semantically
and phonetically close words are likewise close in the induced vector
space since the encoder and decoder RNNs are able to well capture the
acoustic and the contextual relationships at the WSU-level. However, there
is another level of connections that exist more apparently among different
WSUs which the traditional AED models with WSUs output fail to capture - the morphological
relationships.  For example, in addition to the semantic and phonetic
similarity, the words \emph{note, noted, noting, notification, notify,
notified, notifying, notifiable, noticeable, unnoticeable, unnoticeably}
include the same sequence of characters ``not-'', and thus should have
structurally correlated embeddings.  

In a traditional WSU-based AED, the embeddings of the morphologically
related WSUs are initialized and learned independently only through
contextual WSUs and speech in a purely data-driven way. The robust
estimation of so many WSU embeddings (e.g., around 30k) requires a huge amount of
training data. The embeddings are poorly estimated for the WSUs that
rarely occur in the training data.\footnote{For example, in our training set, the WSU ``grandfather's'' occurs only 10 times, but the morphologically similar WSU ``grandfather'' occurs 24,353 times, the WSU ``father'' occurs 371,135 times, the WSU ``grand'' occurs 136,092 times. The embedding of the rare WSU ``grandfather's'' can be more accurately estimated by making use of its morphological relationship with the very frequent WSUs ``grandfather'', ``father'' and ``grand'' through the proposed CA-RNN.}
This is especially problematic for morphologically rich languages, e.g., in
Finnish, a noun has 15 different cases; in French and Spanish, most verbs
have more than 40 inflected forms.

\begin{figure}[htpb!]
	\centering
	\includegraphics[width=0.90\columnwidth]{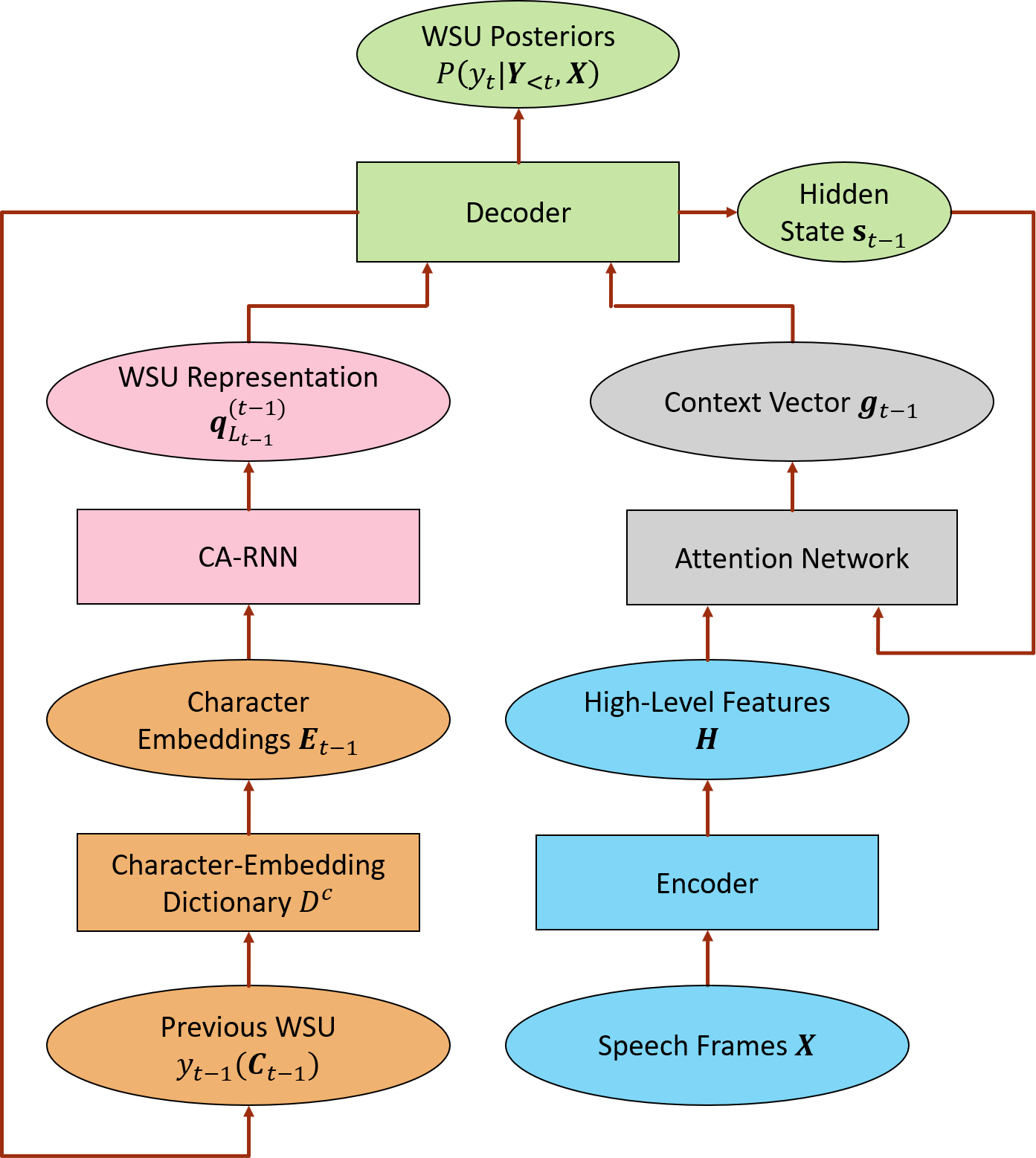}
	\caption{The architecture of CA-AED model for E2E ASR. The
	convolution network generating vector $\mathbf{f}_{t, i}$ is
	omitted for brevity.}
	\label{fig:ca-aed}
\end{figure}

To address this problem, we propose a CA-AED which directly makes use of the
rich morphological relations among WSUs. As shown in Fig. \ref{fig:ca-aed},
based on the existing components of AED, CA-AED introduces an additional
\emph{character-aware (CA) RNN} and replaces the WSU embeddings in
$\mathcal{D}^w$ with WSU representations dynamically generated by this
WSU-independent CA-RNN from \emph{character embeddings}.

\begin{figure}[htpb!]
	\centering
	\includegraphics[width=0.70\columnwidth]{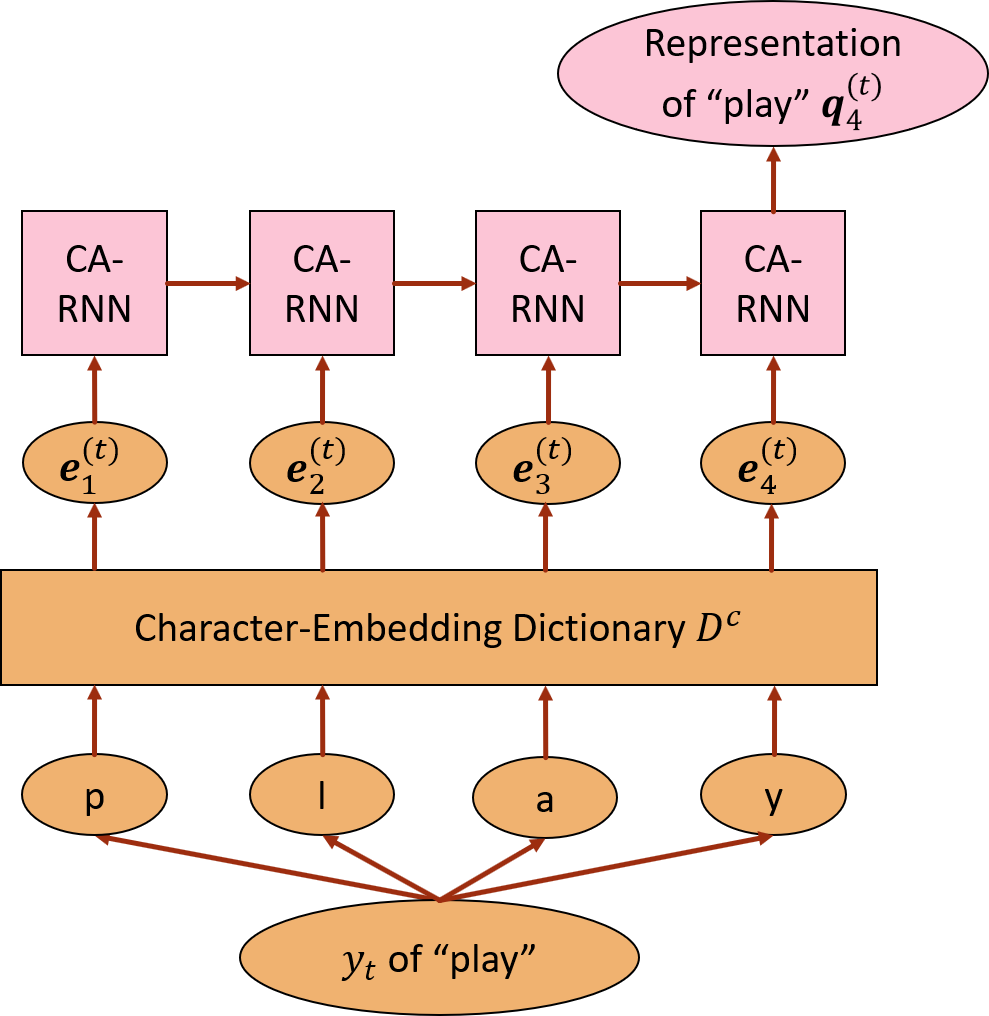}
	\caption{An example of CA-RNN for generating the representation of WSU ``play'' with label $y_t$ from the embeddings of its constituent characters.}
	\label{fig:ca-rnn}
\end{figure}

The WSU $\mathbf{y}_t$ is comprised of a character sequence
$\mathbf{C}_t = \{c_1^{(t)}, \ldots, c_{L_t}^{(t)}\}$, where $L_t$ is length of
$\mathbf{y}_t$ in terms of characters. We construct a character-embedding
dictionary $\mathcal{D}^c$ that maps each character into an embedding
vector. By looking up $\mathcal{D}^c$, we encode $\mathbf{C}_t$ into a sequence
of character embeddings $\mathbf{E}_t = \{\mathbf{e}_1^{(t)}, \ldots,
\mathbf{e}_{L_t}^{(t)}\}$. In CA-AED, the CA-RNN
takes the character-embedding sequence $\mathbf{E}_t$ of the WSU $\mathbf{y}_t$ as the
input and generate a representation for $\mathbf{y}_t$ using its last hidden
state $\mathbf{q}_{L_t}^{(t)}$ as follows.
\begin{align}
	\mathbf{q}_l^{(t)} &= \text{RNN}^{\text{char}}(\mathbf{q}_{l -
		1}^{(t)},
		\mathbf{e}_l^{(t)}), \quad l = 1, \ldots, L_t
\end{align}
$\mathbf{q}_{L_t}^{(t)}$ is then used in place of the WSU embedding
$\mathbf{w}_{t}$ as the input to the decoder RNN below, which further predicts the
conditional probabilities of all possible WSUs via Eq.
\eqref{eqn:decoder_fc}. 
\begin{align}
       & \mathbf{s}_t = \text{RNN}^{\text{dec}}(\mathbf{s}_{t-1}, \mathbf{q}_{L_{t-1}}^{(t-1)} + 
       \mathbf{g}_{t-1}) \label{eqn:decoder_rnn_char}
\end{align}
Fig. \ref{fig:ca-rnn} shows an example of how CA-RNN works.
The encoder and the attention network of CA-AED are
exactly the same as the ones in AED. The character embeddings in
$\mathcal{D}^c$ along with the CA-RNN are jointly trained with the other
parts of CA-AED to minimize cross-entropy loss
$\mathcal{L}_{\text{CE}}$ in Eq. \eqref{eqn:loss}.

With CA-AED, the WSUs sharing the same character substrings are
naturally and explicitly correlated through the CA-RNN so that the
embeddings of rare WSUs can be robustly estimated through
assembling their constituent characters whose embeddings are accurately
learned from abundant training samples. In addition, CA-AED inherits the strong
discriminative power among WSUs by predicting the same set of WSU units at the decoder output layer. 

More importantly, CA-AED entails a much smaller
number of character embeddings (e.g., about 30 in English) and a light-weight CA-RNN to be
learned together with the encoder, decoder and attention network as
opposed to a huge number of WSU embeddings (e.g., about 30k) with 1000 times
more parameters in a conventional AED. Benefiting from modeling the additional morphological
relations, the CA-AED is expected to generate better WSU embeddings for the
decoder and improve the AED-based E2E ASR with significantly reduced
number of parameters. The compression ratio becomes higher
for a CA-AED model of smaller size
since the character embeddings plus CA-RNN save a fixed number of
parameters from WSU embeddings. Therefore, CA-AED has even higher
potential for improving low-footprint AED models on mobile devices.

The training time of CA-AED increases over conventional AED due to the on-the-fly computation of the WSU embeddings from the character embeddings through an LSTM-RNN. But CA-AED saves memory by having smaller number of parameters. During evaluation, CA-AED does not increase the computational cost over the AED model since, before testing, all the WSU embeddings have been pre-computed for once to form the WSU dictionary $\mathcal{D}^w$ by feed the constituent character embeddings of each WSU to the well-trained CA-RNN. Just as a conventional AED model described in Section \ref{sec:aed}, the pre-computed WSU dictionary is then looked up at each decoding step to provide the WSU embedding that the decoder is currently conditioned on to predict the next WSU output.

Note that, during the WSU embedding computation for both training and testing, the CA-RNN resets its memory every time the first character embedding of a WSU is fed as the input. The CA-RNN thus only models the morphology of each WSU, i.e., the statistical relationships among internal characters, without performing any WSU-level language modeling.

\section{Experiments}
We perform E2E ASR using AED and CA-AED with WSUs as the output units on a Microsoft
Windows phone short message dictation (SMD) task.

\subsection{Data Preparation}
The training data consists of 3400 hours of Microsoft internal live US
English Cortana utterances collected through a number of deployed speech
services including voice search and SMD. The test data includes about 5600 utterances (∼6 hours). We explore both the
word pieces and mixed units as the WSUs.  We extract 80-dimensional log Mel
filter bank (LFB) features from the speech signal in both the training and
test set every 10 ms over a 25 ms window.  
We stack 3 consecutive frames and stride the
stacked frame by 30 ms, to form a sequence of 240-dimensional input speech
frames.  We first generate 29190 word pieces as in \cite{sennrich2016neural} and 33755 mixed
units as in \cite{li2018advancing} based on the training transcription and then produce
both word-piece and mixed-unit label sequences serving as the
training targets. We insert a special token \texttt{<space>} in between
every two adjacent words to indicate word boundaries and add tokens
\texttt{<sos>}, \texttt{<eos>} to the beginning and the end of each label
sequence, respectively, to represent sentence boundaries. 

\subsection{AED Baseline System}
We train a baseline WSU-based AED model for E2E ASR as in \cite{meng2019speaker, meng2019domain}. 
The encoder is a bi-directional gated recurrent units (GRU)-RNN
\cite{cho2014properties, chung2014empirical} with 4 or 6 hidden layers, each with 512 hidden units. We use GRU instead of long short-term memory (LSTM) \cite{erdogan2016multi, meng2017deep} for RNN because it has less parameters and is trained faster than LSTM with no loss of performance. Layer normalization \cite{ba2016layer} is applied for each encoder hidden layer. Units
 at the last hidden layer are used as the encoded high-level features. Each WSU is represented
by a 512-dimensional embedding vector in $\mathcal{D}^w$. The decoder is a
uni-directional GRU-RNN with 2 hidden layers, each with 512 hidden units. 
The decoders predicting word pieces and mixed units have 29190
and 33755 output units, respectively.
During training, scheduled sampling \cite{bengio2015scheduled} is applied to the decoder with a sampling probability starting at 0.0 and gradually increasing to 0.4 \cite{chiu2018state}. Dropout \cite{srivastava2014dropout} with a probability of 0.1 is used in both encoder and decoder.
We use 1-D
convolution with a filter size of 15 and 512 output channels to generate
$\mathbf{f}_{t,i}$ and fix $W_h, W_s,
W_f$ as identity matrices to compute the similarity scores $z_{t,i}$ in
Eq. \eqref{eqn:att}.\footnote{We also tried to make $W_h, W_s,
W_f$ full matrices and learn all the parameters in them, but this does not improve the performance. Therefore, we fixed them as identity matrices to save parameters.}  A label-smoothed cross-entropy \cite{chorowski2016towards} loss is minimized during training. Greedy decoding is performed to generate the ASR transcription.\footnote{The WERs can be further reduced by beam search decoding, but greedy decoding is simple with significantly shorter decoding time. The improvement observed in greedy decoding are always consistent with beam search.}
We use PyTorch \cite{paszke2017automatic} for all the experiments.

As shown in Table \ref{table:wer}, AED achieves 9.52\% and 7.75\% WERs with
4-layer and 6-layer encoders, respectively, by predicting word pieces at the output. By predicting mixed-unit output, the WERs decrease to 9.31\% and 7.58\% with 4-layer and 6-layer encoders, respectively. AED achieves better ASR performance with mixed-unit output. We see that mixed-unit AED works slightly better than word-piece AED probably because the former one has more output units than the latter one. But the goal of this paper is not to compare the impact of the two output units on AED performance, but to show the benefits a WSU AED, regardless of the type of WSU, can get from the CA modeling. 

We also trained a character AED by replacing only the WSU output layer of the decoder with 30 units predicting characters. The character AED with 6-layer encoder achieves 9.54\% WER, a 25.9\% relative WER degradation from the mixed-unit AED model \cite{gaur2019acoustic}. Therefore, in this work, we only work on improving our WSU AEDs with CA mechanism.

\begin{table}[h]
\setlength{\tabcolsep}{4.2 pt}
\centering
\begin{tabular}[c]{c|c|c|c|c|c|c}
	\hline
	\hline
	WSU & System & $N_e$ & WER & WERR & $N_p$ & PRR \\
	\hline
	\multirow{4}{*}{\begin{tabular}{@{}c@{}} Word \\ Piece
	\end{tabular}} & 
	\multirow{2}{*}{\begin{tabular}{@{}c@{}} AED
	\end{tabular}} & 4 & 9.52 & - & 44.9 & - \\
        \hhline{~~-----}
	& & 6 & 7.75 & - & 52.2 & - \\
        \hhline{~------}
	& \multirow{2}{*}{\begin{tabular}{@{}c@{}} CA-AED
	\end{tabular}} & 4 & 8.39 & 11.9 & 32.7 & 27.1 \\
        \hhline{~~-----}
	& & 6 & 7.36 & 5.0 & 39.0 & 23.8 \\
	\hline
	\hline
	\multirow{4}{*}{\begin{tabular}{@{}c@{}} Mixed \\ Unit
	\end{tabular}} &
	\multirow{2}{*}{\begin{tabular}{@{}c@{}} AED
	\end{tabular}} & 4 & 9.31 & - & 49.5 & - \\
        \hhline{~~-----}
	& & 6 & 7.58 & - & 55.8 & - \\
        \hhline{~------}
	& \multirow{2}{*}{\begin{tabular}{@{}c@{}} CA-AED
	\end{tabular}} & 4 & 8.52 & 8.5 & 35.0 & 29.3 \\
        \hhline{~~-----}
	& & 6 & 7.35 & 3.0 & 41.3 & 26.0 \\
	\hline
	\hline
\end{tabular}
\vspace{7pt}
\caption{The WER (\%) performance of AED and CA-AED with different WSU output units for E2E ASR on a 3400 hours Microsoft Cortana dataset. $N_e$ is the number of hidden layers in a encoder GRU and $N_p$ (in million) is the total number of model parameters. WERR (\%) and PRR (\%) are the relative WER improvement and the parameter reduction rate of a CA-AED with respect to the AED with the same $N_e$.}
\label{table:wer}
\vspace{-10pt}
\end{table}

\subsection{Character-Aware (CA) AED System}
We further train a CA-AED for E2E ASR with the same training data.
The encoder, decoder and attention network in CA-AED have exactly the same
architectures as the ones in AED.
We map each of the 30 characters into a 256-dimensional embedding vector.
CA-RNN is a GRU with 2 hidden layers and 512 hidden units for each layer. The last state of the top hidden layer of CA-RNN is used as the 
512-dimensional WSU representation.

We vary the number of hidden layers in the encoder $N_e$ to investigate the
effectiveness of CA-AED for different model sizes with different
parameter reduction rates (PRR). As shown in Table \ref{table:wer}, for word-piece model, CA-AED achieves 8.39\% and 7.36\% WERs, respectively, with 4-layer
and 6-layer encoders, which are 11.9\% and 5.0\% relative gains over the AED
baseline system with 27.1\% and 23.8\% less model parameters, respectively.
For mixed-unit model, CA-AED achieves 8.52\% and 7.35\% WERs, respectively,
with 4-layer and 6-layer encoders, which are 8.5\% and 3.0\% relative improved
over the AED baseline system with 29.3\% and 26.0\% reduction in model
parameters, respectively. 

As expected, PRR grows as the number of encoder layers decreases, indicating increased compression ratio.
With a 4-layer encoder, CA-AED performs better for word-piece output, but with a 6-layer encoder, CA-AED achieves similar WERs for mixed-unit and word-piece outputs.
With significantly reduced model parameters, CA-AED improves consistently over AED models
for both word-piece and mixed-unit outputs.  We also observe that the relative WER gain doubles when the encoder downsizes from 6 layers to 4 layers possible because the less accurate acoustic embeddings generated by a weaker encoder of smaller size make more room for the improvement from a more sophisticated WSU representation learned by the CA mechanism. This implies that CA-AED can achieve higher relative improvement 
upon corresponding AED model with a smaller number of parameters, and thus with a higher PRR. Therefore, CA-AED is even more effective in improving the accuracy of low-footprint AED models on mobile devices.

We find that the improved recognition results of CA-AED indeed benefit from the explicit modeling of the morphological relationships among WSUs, such as the relations between the singular and plural forms of a word. From the decoded transcription, we see that CA-AED can accurately recognize the plural form of a word (e.g., sixths, holidays, etc.) during testing given only its singular form (e.g., sixth, holiday, etc.) exists in the training set. However, these unseen plural forms are mistakenly recognized by a conventional AED.

\section{Conclusion}
In this work, we propose a character-aware AED model for
E2E ASR. The CA-AED explicitly models the
morphological relations that exist prevalently among WSUs sharing the same
sequence of characters. An additional CA-RNN is
introduced to generate WSU representations by taking in the embeddings of
their constituent characters. CA-AED makes prediction still at WSU level
while entails only a few character embeddings be learned instead of
a huge set of WSU embeddings. 

Evaluated on a 3400 hours Microsoft Cortana dataset, CA-AED improves the WER
of a traditional AED by up to 11.9\% relatively with 27.1\% fewer parameters with no increase of computational cost during testing. The gain is consistent for both word pieces or mixed units as the output units.
CA-AED has great potential in improving small-footprint model on mobile
devices, as the relative gain is higher over the AED models with fewer
parameters. 

\vfill\pagebreak

\bibliographystyle{IEEEtran}
\bibliography{refs}

\end{document}